\documentstyle[aps,prl,floats,twocolumn,epsf]{revtex}

\newcommand{\xdir}{$[ 1\overline{1} 0 ]$}
\newcommand{\ydir}{$[ 110 ]$}
\newcommand{\zdir}{$[ 001 ]$}
\newcommand{\xydir}{$[ 100 ]$}
\newcommand{\yxdir}{$[ 010 ]$}
\newcommand{\Bpar}{$B_{\parallel}$}
\newcommand{\Bparcrit}{$B_{\parallel}^{*}$}

\begin{document}
\draft

\wideabs{
\title{An Investigation of Orientational Symmetry-Breaking
Mechanisms in High Landau Levels}

\author{K.~B.~Cooper$^1$, M.~P. Lilly$^{1}$\cite{lillyaddress}, J.~P. Eisenstein$^1$, 
T.~Jungwirth$^{2,3}$, L.~N. Pfeiffer$^4$, and K. W. West$^4$}

\address{$^1$California Institute of Technology, Pasadena, CA 91125 \\
	 $^2$University of Texas, Austin, TX 78712\\        
	 $^3$Institute of Physics ASCR, Cukrovarnick\'a 10, 162 53 Praha 6, Czech Republic\\
	 $^4$Bell Laboratories, Lucent Technologies, Murray Hill, NJ 07974\\}

\date{\today}

\maketitle

\begin{abstract} The principal axes of the recently discovered anisotropic 
phases of 2D electron systems at high Landau level occupancy are consistently 
oriented relative to the crystal axes of the host semiconductor.  The 
nature of the native rotational symmetry breaking field responsible for this
preferential orientation remains unknown. Here we report on experiments designed 
to investigate the origin and magnitude of this symmetry breaking field.  Our 
results suggest that neither micron-scale surface roughness features nor the 
precise symmetry of the quantum well potential confining the 2D system are 
important factors.  By combining tilted field transport measurements with 
detailed self-consistent calculations we estimate that the native anisotropy 
energy, whatever its origin, is typically $\sim$~1 mK per electron.

\end{abstract}

}

A fundamental goal of contemporary condensed matter physics is to
understand the ground state of the two-dimensional electron
system (2DES) and to explore its experimental signatures. In the
presence of a large perpendicular magnetic field $B$, the kinetic energy
of 2D electrons becomes quantized into discrete massively
degenerate Landau levels (LLs).  At high $B$ only the lowest ($N$ = 0) LL is 
occupied, and the 2DES exhibits its most spectacular phenomenon: the fractional 
quantized Hall effect (FQHE)\cite{review}. 
At lower magnetic fields the higher LLs become occupied, and in the third or
higher ($N$~$\ge$~2) LL no FQHE states have been observed.
Recent experiments have
nevertheless uncovered extraordinary transport signatures unique to 
the high LL regime, pointing to the existence of a new class of many-body states 
distinct from the incompressible quantum fluids responsible for the 
FQHE \cite{lilly1,du}.  The most striking of these signatures is the rapid 
development, at very low temperatures, of strong anisotropies in the 
longitudinal resistance of the 2D electron system.

Observed only in very high mobility samples and at temperatures below about 
100 mK, the anisotropies in the longitudinal resistance are strongest near 
half-filling of the $N$~=~2 and several higher Landau levels.  This corresponds to 
Landau level filling fractions $\nu$, defined as the ratio of the electron 
density $N_s$ to the degeneracy $eB/h$ of a single spin-resolved LL, of
$\nu$~=~9/2, 11/2, 13/2, etc.  For 2D electron systems in GaAs/AlGaAs 
heterostructures grown on \zdir-oriented GaAs substrates, the anisotropies are 
consistently disposed so that the ``hard'' transport direction is parallel to 
the \xdir~crystallographic direction while the ``easy'' direction is parallel to 
\ydir.  Although a persuasive picture of how anisotropic electronic ground 
states develop at high Landau level occupancy now exists, there is still no 
understanding of why they are consistently oriented relative to the host crystal 
axes.  The origin and magnitude of the necessary native symmetry breaking field is 
the focus of this paper.

The resistance anisotropy near half-filling of high LLs has been widely 
interpreted as evidence for charge-density-wave (CDW) ground states. Hartree-Fock
(HF) calculations\cite{KFS,MC} have suggested that 2D electrons in half-filled
high LLs form a unidirectional CDW, or ``stripe'' phase.  At $\nu$~=~9/2, 
for example, the system is expected to phase separate into alternating regions 
of $\nu$~=~4 and $\nu$~=~5 with a period of about three times the cyclotron radius 
$R_c$~=~$\hbar k_F /eB$, or about 100 nm in typical samples.  If these stripes are 
somehow preferentially oriented, anisotropy in the longitudinal 
resistance of the sample would likely result.  Transport currents flowing 
perpendicular to such stripes would presumably encounter greater resistance than 
currents flowing parallel to them.

More recent theoretical work has generally supported the early HF predictions, 
albeit with some significant modifications.  Numerical exact diagonalizations of 
small systems of electrons have uncovered sharp features in the susceptibility and 
structure factor of the 2DES in high LLs consistent with stripe formation at 
half filling\cite{rezayi}.  Refinements to the HF approach suggest that the 
stripes are themselves unstable against modulations along their 
length\cite{fertig,macdonald}.  The resulting ``smectic crystal'' would 
presumably be pinned and therefore insulating at very low temperatures.  Quantum 
and/or thermal fluctuations are expected to melt the system and render it 
analogous to a nematic liquid crystal\cite{fradkin}.  A nematic has 
translational but not rotational symmetry and might therefore exhibit 
anisotropic transport if a preferred direction exists.  At still higher 
temperatures a nematic to isotropic transition is predicted to restore 
rotational symmetry\cite{fradkin,wexler}.  

The consistency of the orientation of the transport anisotropy in high LLs has 
been established through experiments on a large number of GaAs/AlGaAs 
heterostructure samples.  Grown by molecular beam epitaxy (MBE), these 
modulation-doped 2D electron systems have low temperature mobilities typically 
around $\rm 10^7~cm^2/V~s$ and densities ranging from 
$\rm 1.5-2.7\times10^{11}~cm^{-2}$.  
The samples are usually illuminated briefly at low temperature with a red light 
emitting diode to enhance their transport characteristics but, as explained 
below, the existence and orientation of the high Landau level resistance 
anisotropy does not depend upon this.  Although most measurements have been made 
using square samples whose sides are parallel to the \xdir ~and \ydir 
~directions, specific additional experiments have shown that the orientation of 
the transport anisotropy is insensitive to the geometry of the conducting 
region, at least in mm-sized samples\cite{lilly2}.  

In this paper we report new results which narrow down the list of possible 
native symmetry breaking effects that might orient the high LL anisotropy.  
Specifically, we show that neither the micron-scale morphology of the sample surface, as 
revealed by atomic force microscopy (AFM), nor the shape of the heterojunction 
potential confining the electron gas to two dimensions appear to be significant 
factors.  In addition, by combining experimental results obtained using tilted 
magnetic fields with detailed numerical calculations, we estimate the strength 
of the still unknown symmetry breaking field.  

Data from three samples (A, B, and C) will be presented here.  Sample A consists 
of a single 30 nm GaAs quantum well embedded in thick layers of AlGaAs.  Doping 
is provided by Si layers positioned symmetrically above and below the quantum 
well.  Before illumination the 2DES in this sample has a density of 
$\rm 2.74\times10^{11}~cm^{-2}$ and a low temperature mobility of 
$\rm 1.8\times10^7~cm^2/V~s$.  Brief illumination at 1.6 K changes these values to 
$\rm 2.54\times10^{11}~cm^{-2}$ and $\rm 2.1\times10^7~cm^2/V~s$, respectively.  
Samples B and C are conventional single heterointerfaces.  After illumination 
these samples respectively have densities of 2.67 and $\rm 
1.48\times10^{11}~cm^{-2}$ and mobilities of 0.9 and $\rm 1.1\times10^7~cm^2/V~s$. 
Each sample was 
cleaved from its parent MBE wafer into a square, 3 or 5 mm on a side, and eight 
diffused In 
contacts were positioned at the corners and midpoints of the sides.  Transport 
measurements were made with an excitation current of $I$~$\leq$~20 nA at 13~Hz 
driven between midpoint contacts on opposite sides of the sample. Longitudinal 
resistance data were obtained by measuring the resulting voltage drop between 
corner contacts. Each of these samples clearly exhibits the characteristic low 
temperature resistance anisotropies near half-filling of the $N$~$\geq$~2 LLs first 
reported by Lilly, {\it et al.}\cite{lilly1}.

The inevitable imperfections which arise during crystal growth are natural 
candidates for an extrinsic mechanism capable of orienting the anisotropic 
electronic phases in high LLs.  Since the kinetics of MBE growth is known to be 
anisotropic, it is easy to imagine that these imperfections might be 
preferentially oriented.  This would break the rotational symmetry of the local 
environment of the 2DES and possibly influence its transport properties.  As it 
is obviously difficult to directly observe the local environment of the 2DES, 
which is buried inside the sample, we have resorted instead to an
examination of the exposed top surface of the heterostructure via 
AFM\cite{afminfo}.  
Although this surface is typically 150 nm above the plane of the 2DES, the 
predicted wavelength of the CDW in high LLs is of comparable magnitude.

Nine high mobility GaAs heterostructure wafers (including those of samples A, B, and C) were 
examined in search of anisotropic surface morphologies.  Each contained a 2DES 
with mobility exceeding $\rm 0.8\times10^7~cm^2/V~s$ and density in the range
of $\rm1.5-2.7\times10^{11}~cm^{-2}$.  Unambiguous determinations of the 
\xdir~and \ydir~directions were made using one or more 
techniques\cite{techniques}. Samples from all nine wafers showed strong resistance 
anisotropies at $\nu$~=~9/2, 11/2, etc.\ with the high resistance direction along 
\xdir.   

\begin{figure}[h]
\begin{center}
\epsfxsize=3.3in
\epsffile[18 18 291 400]{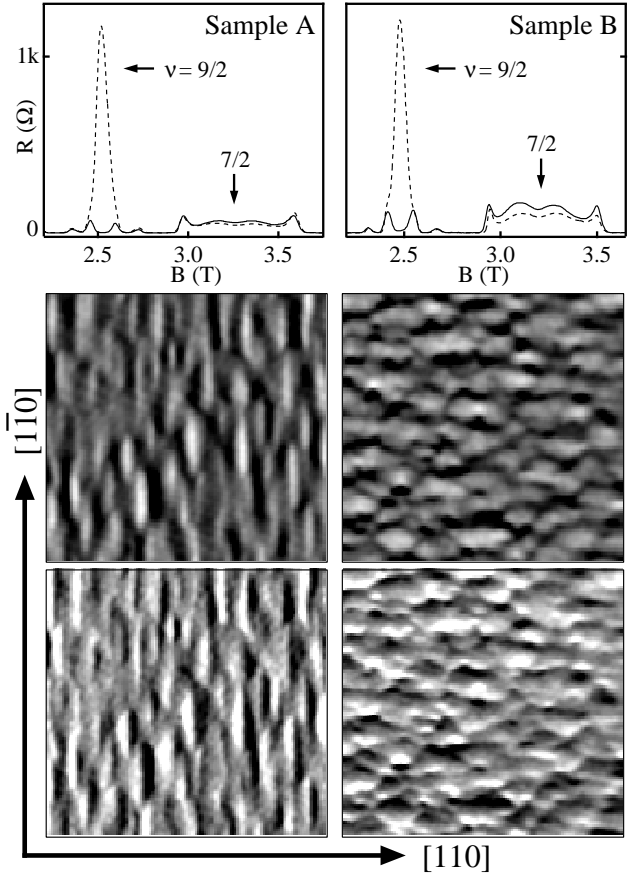}
\end{center}
\caption[Figure 1]{
Top panels: Longitudinal resistance data taken at 50 mK in samples A and B.  
Dashed curves: resistance along \xdir; solid curves: resistance along \ydir. 
Strongly anisotropic transport is evident around $\nu$~=~9/2.  
Middle panels: 
Height AFM images ($16 \times 16~\mu$m) of the samples' parent wafers. 
The prominent mounds are $5-10$~nm tall.
Bottom panels: 
Corresponding amplitude AFM images. While for both 
samples the hard and easy transport directions at $\nu$~=~9/2 lie along \xdir\ and 
\ydir, respectively, the orientations of their surface morphologies 
are perpendicular to one another. 
}
\end{figure}

Figure 1 shows the results of transport and AFM studies of samples A and B.
The top two panels show longitudinal resistance data, taken at $T$~=~50~mK, from 
the filling factor range 3~$<$~$\nu$~$<$~5.  Near $\nu$~=~9/2 the strong resistance 
anisotropy characteristic of half-filled high Landau levels is clearly evident. 
As the data show, the ``hard'' transport direction is along \xdir.  In contrast, 
the resistance near $\nu$~=~7/2 is nearly isotropic.  This filling factor, which 
lies in the $N$~=~1 LL, supports a weak fractional quantized Hall state which gives 
way to an anisotropic state only when a large magnetic field component parallel 
to the 2D plane is applied\cite{lilly2,pan}.  It is worth noting that the data from sample A 
shown here was obtained without illumination, verifying that the anisotropy 
effect does not depend upon this.  Data from sample A after illumination will be
discussed below.

The lower panels of Fig.~1 display $16 \times 16~\mu$m AFM images
of the \zdir-oriented surfaces of the wafers from which samples A and B
were cleaved, and 
the \xdir\ and \ydir\ directions are indicated by arrows.  
The images were obtained at room
temperature using the ``tapping mode" technique.
In tapping mode, the microscope's cantilever tip oscillates at its
resonant frequency as it is scanned across a surface. A feedback
loop attempts to maintain the tip oscillation at a fixed amplitude by
adjusting the average height of the tip above the sample surface; this
average tip displacement is recorded as ``height data." Transient
changes in the oscillation amplitude, roughly a measure of the height
derivative in the scan direction, are monitored as ``amplitude data."
The middle panels of Fig. 1 display height data,
while the bottom panels display the corresponding amplitude data. 

The inevitable slow drifts of the cantilever tip during image acquisition,
while leaving the amplitude data unaffected, can
obscure shallow
surface features in the height data.
To compensate for these drifts,
the height images in the middle panels of Fig.~1 were subjected to
a high-pass filter in the scan direction.
Because this filtering risks a loss of information of surface height
variations perpendicular to the scan direction,
four independent scans parallel to \xdir, \ydir, \yxdir, and
\xydir\
were performed for each surface. These results (not shown here) confirm that the images 
of Fig.~1 faithfully represent all
relevant morphology and that no spurious features related to the scan
direction are present.  Furthermore, the characteristic surface features revealed in Fig.~1 
were found to cover each wafer's entire surface, indicating that their occurrence is a generic
property of the MBE growth process.

Both surfaces in Fig.~1 exhibit shallow ($\leq10$ nm), anisotropic surface features
on the micron length scale.  However, two important differences should be noted:
first, the orientation of the surface features of the two samples are \emph{orthogonal} 
to each other, and second, the general shapes of the surface features differ
for the two samples.  The surface of sample A, for example, is covered by long and straight
cigar-shaped mounds oriented with their major axes along \xdir.  Sample B's surface,
on the other hand, exhibits more irregularly-shaped mounds clearly elongated along \ydir\ and  separated by narrow,
meandering valleys.

These results are typical of the variety of surface features seen on seven other wafers examined,
where AFM images revealed varying degrees of mounding.
Three of the seven showed 
mounds elongated along \ydir, two showed rounded mounds with virtually 
no preferred direction, and two showed mounds elongated along 
\xdir.  We emphasize, however, that in all the heterostructures the hard and easy axes of 
the resistance anisotropy in high LLs are along \xdir\ and \ydir, respectively.  
Although the ``quality'' of the transport data does vary among the samples, it 
is sensibly correlated with the mobility of the 2DES and shows no evident 
dependence on the surface morphology.

The presence of anisotropic mounding, or roughness, on thin GaAs epilayers grown by MBE
on \zdir-oriented GaAs substrates has been noted previously by Orme,
{\it et al.}\cite{orr}. The competition between step-flow growth and
island nucleation, in conjunction with anisotropic surface diffusion,
was found to result in micron-sized mounds elongated along \xdir.
Although Orme, {\it et al.} did not observe mounding oriented along
\ydir, as we find in 4 of 9 samples studied, we note that our
heterostructures are much thicker and more complex (consisting of
hundreds of layers of GaAs and AlGaAs) than simple GaAs epilayers. As
Orme, {\it et al.} found the tendency to mound to be very sensitive to
growth temperature, substrate miscut angle, and epilayer thickness, the
greater diversity of our results seems less surprising. 

The lack of a consistent crystallographic orientation of the
micron-scale surface roughness features on our samples contrasts sharply
with the highly consistent alignment of the transport anisotropy axes
exhibited by the 2DES at $\nu$~=~9/2, 11/2, etc., in the same samples.
From this we conclude that micron-scale surface roughness does not
reflect the symmetry breaking field responsible for the alignment of the
anisotropic electronic phases in high LLs. Whether finer scale surface
features or unseen subsurface defects are involved remains to be
determined. While we agree with Willett, {\it et al.}\cite{willett},
that {\it sufficiently severe} surface roughness (whether as-grown or
externally imposed) can influence 2DES transport, we do not find
that the orientation of micron-scale surface features is
always correlated with transport anisotropies in high LLs. 

Another effect which might break the symmetry between the \xdir\ and \ydir\ 
directions in a 2DES on a \zdir-oriented surface is related to the shape of the 
potential well confining the electrons.  Kroemer\cite{kroemer} has noted that 
while transport in {\it bulk} GaAs is invariant under $\pi$/2 rotations 
about the \zdir\ axis, this may not be the case in a 2DES.  If the potential 
confining the electrons to the \zdir\ plane lacks inversion symmetry the 
transport coefficients need only be invariant under $\pi$ rotations.  While it 
seems unlikely that the band structure effects resulting from this lack of 
inversion symmetry could explain the large magnitude and sharp temperature and 
filling factor dependences of the high LL transport anisotropy, they may be 
sufficient to orient an otherwise spontaneously generated anisotropic ground 
state.  Rosenow, {\it et al.}\cite{rosenow} and Takhtamirov, {\it et al.}\cite{takh} 
have stressed this point and have 
estimated that the reduced symmetry of typical single-interface heterojunctions 
leads to small ($\sim$~0.1\%) anisotropies in the 2D electron effective mass. 

The transport measurements of samples A and B in Fig.~1 allow for examination of 
this idea. Sample A consists of a single 30 nm GaAs quantum well embedded within 
$\rm{Al_{0.24}Ga_{0.76}As}$.  Silicon doping layers positioned 98 nm above and 
below the quantum well produce a symmetrically confined 2DES. Sample B, on the 
other hand, contains a single GaAs/AlGaAs interface doped from one side only.  
The 2DES in this sample is therefore confined by an asymmetric, roughly 
triangular, potential.  In spite of this structural difference the two samples 
show qualitatively identical high LL transport features.  In particular, the 
orientation of the anisotropy axes relative to the crystal axes is the same in 
the two samples.  While one might argue that uncontrolled microscopic 
irregularities would destroy the symmetry of square quantum wells, it has been 
convincingly demonstrated, in 2D hole systems, that the symmetry of the 
confinement potential can have readily measurable transport consequences under 
appropriate circumstances\cite{jpeholes,shayegan}.  Therefore, the lack of any 
significant differences in the high LL transport in samples A and B suggests 
that the symmetry of the confinement potential is also not a major factor in 
determining the orientation of the underlying anisotropic electronic states.
 
The transport data from sample A displayed in Fig. 1, which were obtained 
without prior low temperature illumination, demonstrate that such illumination 
is not required to observe anisotropic transport in high Landau levels.   
Illumination is usually employed because it has been found empirically to often 
improve the quality of high field transport data.  The precise way in which this 
occurs is not well understood, but in addition to typically increasing the 2D 
density and mobility, illumination also seems to improve the homogeneity of the 
electron gas. In any case, Fig. 2 demonstrates this improvement in 
sample A.  The minima in the resistance measured along \ydir\ at $\nu$~=~9/2 and 
11/2 are much deeper, the isotropic re-entrant integer quantized Hall states in 
the flanks of the Landau levels are much better 
developed\cite{lilly1,du,cooper}, and additional structure is 
evident.  Nonetheless, these data show clearly that illumination is not an 
important factor in the origin or orientation of the anisotropic phases of 2D 
electrons in high Landau levels.

\begin{figure}[ht]
\begin{center}
\epsfxsize=3.3in
\epsffile[10 10 400 400]{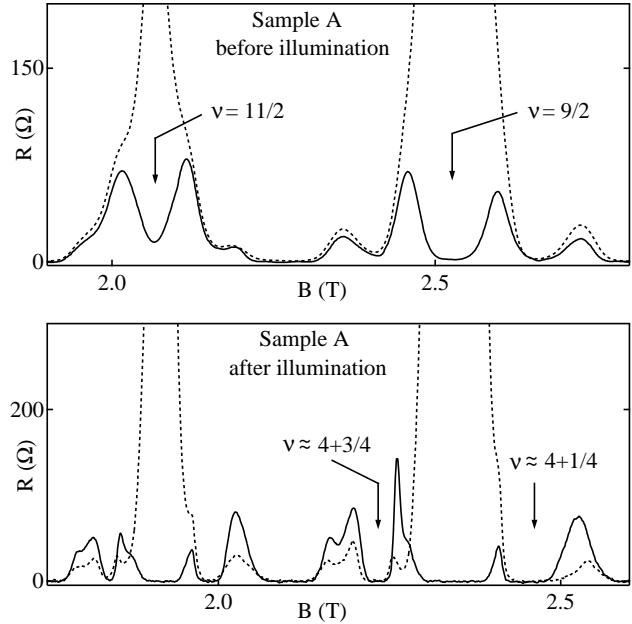}
\end{center}
\caption[figure 2]{
Effect of low-temperature illumination by a red light emitting diode on the
transport characteristics of sample A at 50 mK.  Top: Longitudinal resistance
along \xdir~(dashed) and \ydir~(solid) before illumination in the region
$4 \le \nu \le 6$.  Bottom: The same measurement after illumination.  Note 
the improvement in the degree of anisotropy at 1/2 filling and the increased prominence
of the re-entrant integer states near 1/4 and 3/4 filling.
}
\end{figure}

A magnetic field component \Bpar\ applied parallel to the plane of the
2DES provides an important tool for investigating the magnitude of the
native symmetry breaking mechanism. Such an in-plane field allows the
controlled introduction of an {\it additional} symmetry breaker which is
known to be capable of altering the high Landau level transport
anisotropy axes \cite{lilly2,pan}. For example, only a relatively small
\Bpar\ pointed along \ydir\ is required to \emph{interchange} the
original principal axes of anisotropy, with the new easy and hard
directions along \xdir\ and \ydir\ respectively. Prior to this
interchange, 
there is a special value of the in-plane field, \Bparcrit, where the
transport in the sample becomes approximately isotropic.
Remarkably, if the in-plane field is instead directed along \xdir,
the principal axes of anisotropic transport do not reverse.

Hartree-Fock calculations within the undirectional CDW model can account for these 
observations: the coupling between \Bpar\ and the finite thickness of the 2D 
layer favors stripes aligned perpendicular to \Bpar\ under typical 
circumstances\cite{jungwirth1,phillips}.  Assuming that the high 
resistance direction is perpendicular to the stripes, this conclusion is in 
agreement with experiment. Similar calculations have also been successful in 
explaining the more complex anisotropic transport behavior exhibited at high 
\Bpar\ by quasi-2D systems in wide quantum wells having two occupied 
subbands\cite{jungwirth2}. These theoretical analyses also yield quantitative 
estimates of the field-induced anisotropy energy, $E_A$, defined as the difference in 
energy (per electron) of undirectional CDWs aligned parallel and perpendicular 
to the in-plane magnetic field.  
By evaluating the anisotropy energy at \Bparcrit, these calculations provide
an estimate of the magnitude of the native 
symmetry breaking field which orients the transport anisotropy in the absense
of an in-plane field.  

Figure 3 shows the effect of an in-plane magnetic field directed along \ydir\ on 
transport in the vicinity of $\nu=9/2$ in samples A and C (data from sample B 
has been published previously\cite{lilly2}).  These data, which are obtained by 
tilting the sample relative to the axis of a single superconducting solenoid, 
clearly reveal the interchange of the anisotropy axes. For sample A a tilt of 
only $\theta$~=~$6^{\circ}$ is sufficient to render the transport roughly 
isotropic.  This corresponds to \Bparcrit~$\approx$~0.24 T at $\nu$~=~9/2. By 
$\theta$~=~$10^{\circ}$ the transport anisotropy is fully restored, albeit with 
interchanged axes.  Samples B and C exhibit switching fields of approximately 
\Bparcrit~$\approx$~0.50 and 0.55 T respectively.
  
\begin{figure}[h] 
\begin{center} 
\epsfxsize=3.3in 
\epsffile[6 0 470 470]{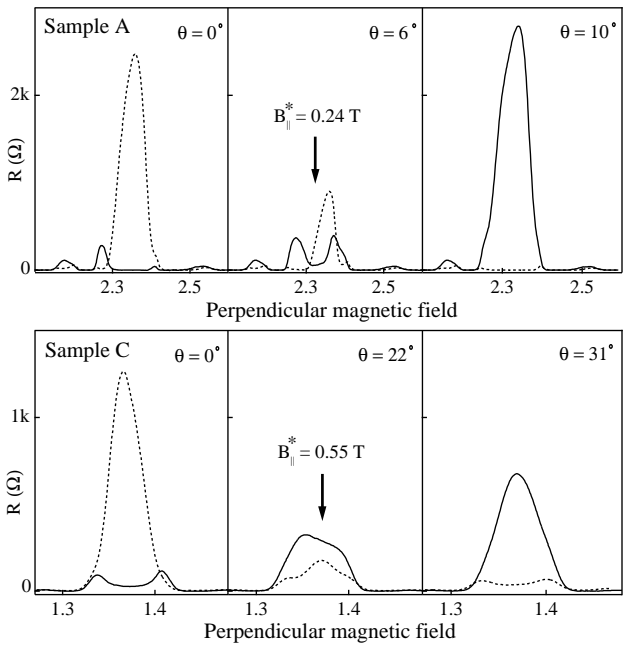} 
\end{center} 
\caption[figure 3]
{ Effect of a parallel
magnetic field in the \ydir\ direction on anisotropic transport at
$\nu$~=~9/2 of samples A (50 mK) and C (25 mK). Tilting about the \xdir\
axis results in a reversal of the principal axes of anisotropy. Dashed:
longitudinal resistance along \xdir . Solid: longitudinal resistance along \ydir .} 
\end{figure}

Theoretical field-induced anisotropy energies are plotted in Figure 4.
The many-body RPA/Hartree-Fock calculations are combined with a self-consistent 
local-density-approximation description of one-particle
states\cite{jungwirth1}, obtained numerically for the specific
structural parameters of samples A, B, and C.  In agreement with the
experimental observations, the unidirectional CDWs prefer being aligned
perpendicular to the in-plane magnetic field direction ($E_A \ge 0$).
The agreement between experiment and theory on this point offers strong
support to the overall CDW picture of the anisotropic phases in high
Landau levels.  After all, this theoretical result is not obvious
\emph{a priori}, as the interaction between a parallel magnetic field
and the 2DES depends sensitively on the detailed finite-width profile of
the 2DES and the nature of screening in high magnetic fields.  Indeed,
previous calculations have shown that unlike strongly confined 2D
electron systems where the CDW orients perpendicular to
\Bpar\cite{jungwirth1}, in two-subband quantum wells the preferred CDW
orientation is \emph{parallel} to the in-plane field\cite{jungwirth2}.  This
observation helps explain the remarkable difference in the dependence of
$E_A$ on \Bpar\ in samples A and B despite the samples having nearly
identical 2D electron densities.  In sample B, the unoccupied second
subband is much closer to the chemical potential than in sample A.
Thus, being more similar to a system with two subbands occupied, a CDW in
sample B should gain a smaller energy advantage by aligning
perpendicular to a given in-plane field.

\begin{figure}[h] 
\begin{center} 
\epsfxsize=3.3in 
\epsffile[0 0 335 335]{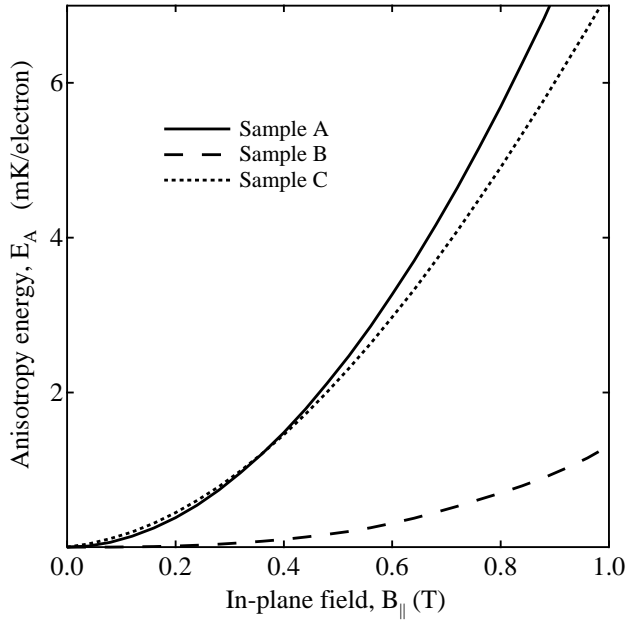} 
\end{center} 
\caption[figure 4]
{ Theoretical results of the anisotropy
energy versus an in-plane magnetic field for a unidirectional CDW at $\nu$~=~9/2 in samples
A, B, and C.  The CDW periods in the three samples, at \Bpar~=~0, were calculated to be 
108, 106, and 143 nm respectively.
Evaluated at the switching field \Bparcrit, the anisotropy energy
yields an estimate of the native pinning energy of a CDW phase.
} 
\end{figure}

The calculated field-induced anisotropy energies at the measured $\nu$~=~9/2
switching fields \Bparcrit\ are 0.5, 0.2, and 2.4 mK per electron
in samples A, B, and C respectively. These energies are far smaller than
the $\sim$~100 mK temperature scale that apparently governs the formation
of the anisotropic electronic phase (i.e.\ the temperature at which
transport becomes anisotropic). This disparity supports the model of a
robust CDW oriented by a relatively weak native symmetry breaking field.
Although the origin of the native symmetry breaking field remains
unknown, these estimates of the anisotropy energy should offer useful
constraints on proposed mechanisms. 

The experimental results reported here suggest that two possible
symmetry breaking effects, anisotropic surface roughness and
confinement potential geometry, are not important factors in
determining the orientation of the newly discovered anisotropic phases
of 2D electrons in high Landau levels. Nevertheless, other possibilities
remain. For example, Fil\cite{fil} has argued that the
piezoelectricity of GaAs leads to lower energy for CDW's approximately aligned along
either \xdir\ or \ydir. Although this does not explain what further
distinguishes between these two directions, it does establish a
mechanism which breaks rotational symmetry and favors directions which
do have relevance to experiment. 

In conclusion, we have addressed the still unresolved issue of how the
anisotropic electronic states in high LLs become oriented relative to the 
crystallographic directions of the host semiconductor lattice. We have found 
that neither micron-scale surface roughness nor the symmetry of the confinement 
potential are major factors. In conjunction with detailed theoretical 
calculations, tilted-field experiments support the overall charge density wave 
picture of the ground state of 2D electrons in high Landau levels and yield 
quantitative estimates of the native anisotropy energy for three different 
samples. 

We thank M. Roukes for his generous loan of the atomic force microscope
used in this work, E. Rashba for
pointing out ref.\ 26 to us, R. Willett for showing us his data prior to
publication, and S. Das Sarma and
A.C. Gossard for helpful discussions.  This work is supported by the NSF
under grant DMR-0070890, the
DOE under grant DE-FG03-99ER45766, and the Grant Agency of the Czech Republic under grant
202/01/0754.


\begin{references}
\bibitem{techniques}The \ydir\ and \xdir\ directions were distinguished via 
wet-etch profiling, observation of ``bowtie'' defects on the sample surface, and 
knowledge of the parent wafer orientation. No conflicts arose.
\end{references}
\end{document}